# ROLE OF HYDRODYNAMIC DRAINAGE IN ADHESION TO WET SURFACES

*Rohini Gupta and Joelle Frechette*
*Chemical and Biomolecular Engineering Department, Johns Hopkins University*
*Baltimore, MD, 21218, USA*
*Email: jfrechette@jhu.edu*

## Introduction

Biological creatures rely on a combination of specific interactions (such as ligand-binding) and weaker non-specific interactions (van der Waals, capillarity, electrostatics, hydrodynamics) for adhesion, climbing, and gripping. Perhaps the most studied case is that of the gecko being able to climb on almost any surfaces without the use of a mediating fluid (such as mucus) and leveraging weak van der Waals forces.[1-6] Geckos are able to climb up a smooth wall due to the formation of small inhomogeneous contacts with the wall surface [7]. The role of van der Waals forces, however, is expected to be significantly reduced for the case of adhesion mediated by a viscous fluid (wet or flooded adhesion), as is the case for tree frogs. Viscous and capillary forces should also play an important role in adhesion when a fluid is present, whereas they are completely absent in dry adhesion.[8,9] Recent experiments with tree frogs have started to shine light on this phenomenon, however, they have also raised interesting questions about the adhesion mechanisms.[10] The problem of wet adhesion is especially challenging because of the added complexity associated with the interpretation of dynamic viscous forces (i.e. their dependence on approach and retraction velocities). This poster present our preliminary results towards the investigation of the role of surface structure, and more specifically that of surface channels, on hydrodynamic forces.

## Experimental

The SFA [11,12] was employed to measure forces between two surfaces as a function of their separation under a constant drive velocity. The normal forces were measured with leaf springs at a resolution down to 10nN [13]. The normal separation between surfaces is varied with 0.1nm resolution using microstepping motors and measured independently using multiple beam interferometry (MBI) [14]. The drive velocity was varied via the microstepping motor to perform experiments in a quasi-static regime (velocities less than 1 nm/s) to about 1 micron/s (which is very fast considering that the radius of curvature of between 1-2 cm).

The force curves were measured between a spin-coated PMMA film and a silver surface. The surfaces of interest were submerged in a 48cSt silicon oil. The force curves were analyzed following Reynolds' continuum treatment whereas the force is obtained from the integration of the Navier-Stokes equation with the no slip boundary condition for an incompressible fluid in the lubrication limit. [15] The force were calculated numerically using a fourth order Runge-Kutta method (using ode45 in Matlab).[16]

## Results and Discussion

Using the SFA we measured the hydrodynamic drainage forces between a smooth 6μm thick PMMA film and a silver surface immersed in a 48cSt silicon oil. The measured repulsive force upon bringing the surfaces together is shown in Fig. 1 for driving velocities that span 4 orders of magnitude (from 0.6 nm/s up to 700 nm/s). The solid lines shows theoretical predictions calculated from solving the Reynold's equation. As seen in Fig. 1, we obtain good agreement between our experiments and theoretical predictions. The surface wall, however, had to be shifted by 2-8 nm to account for adsorption of the oil on the surfaces. As expected from lubrication theory, the magnitude and range of the repulsive barrier depends on the drive velocity consistent with the idea that a rate-dependent repulsive barrier has to be overcome to drain the fluid in the gap and make good contact between the interacting surfaces.

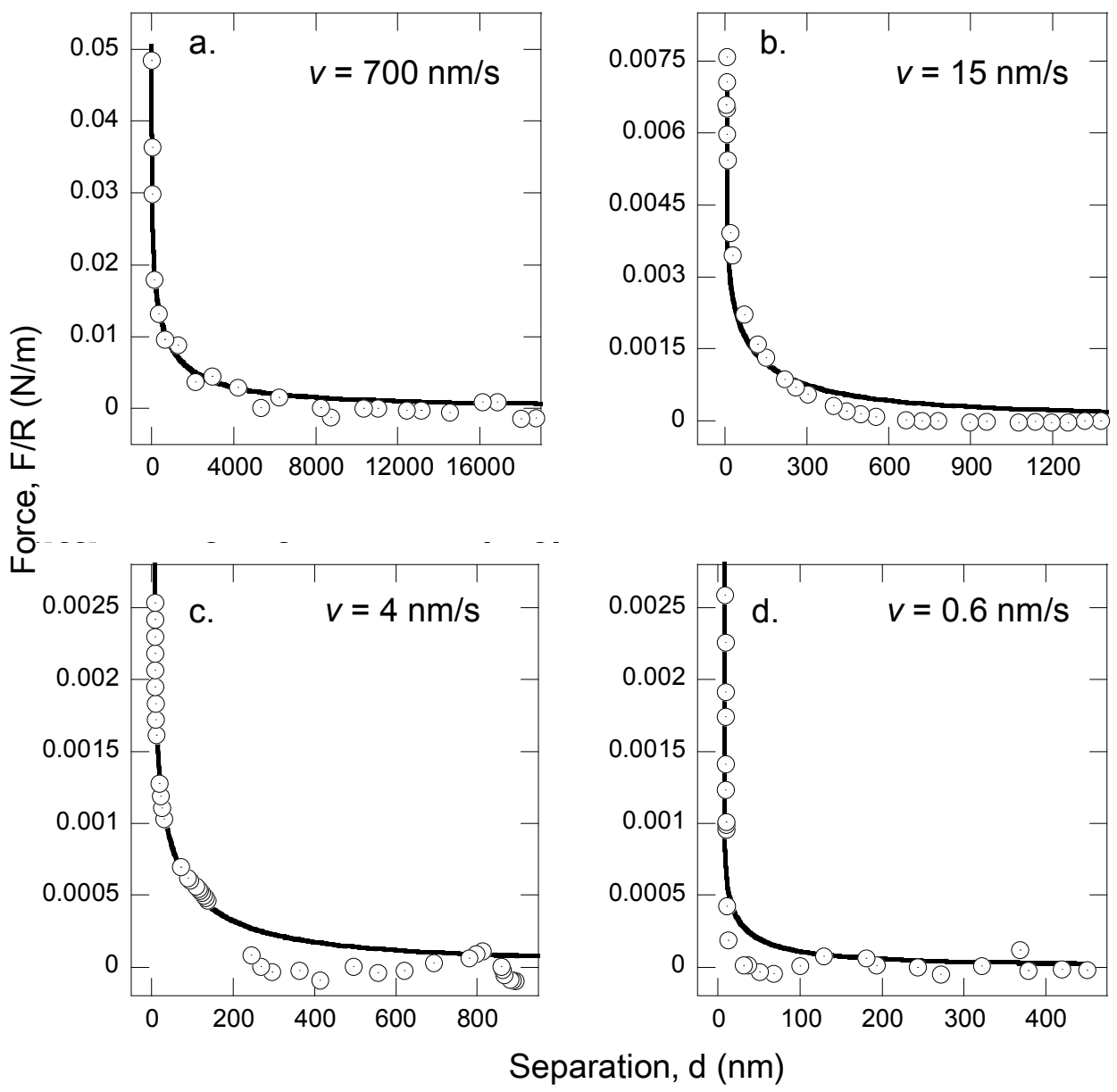

Figure 1. Drainage force required to bring a smooth PMMA surface in contact with a silver surface measured at 4 different driving velocities.

Adhesion forces were also measured when the surfaces were pulled apart (see Fig. 2) and good agreement with theory is observed. As expected, the adhesion force scales with the pull-out velocity (An adhesive force of approximately 30mN/m is obtained for a drive velocity of 700 nm/s and an adhesive force of 0.3mN/m is obtained for a drive velocity of 0.6 nm/s). Measurement of drainage forces involving a surface patterned with squeeze out channels will be performed and the results will be compared to those obtained for smooth surfaces.

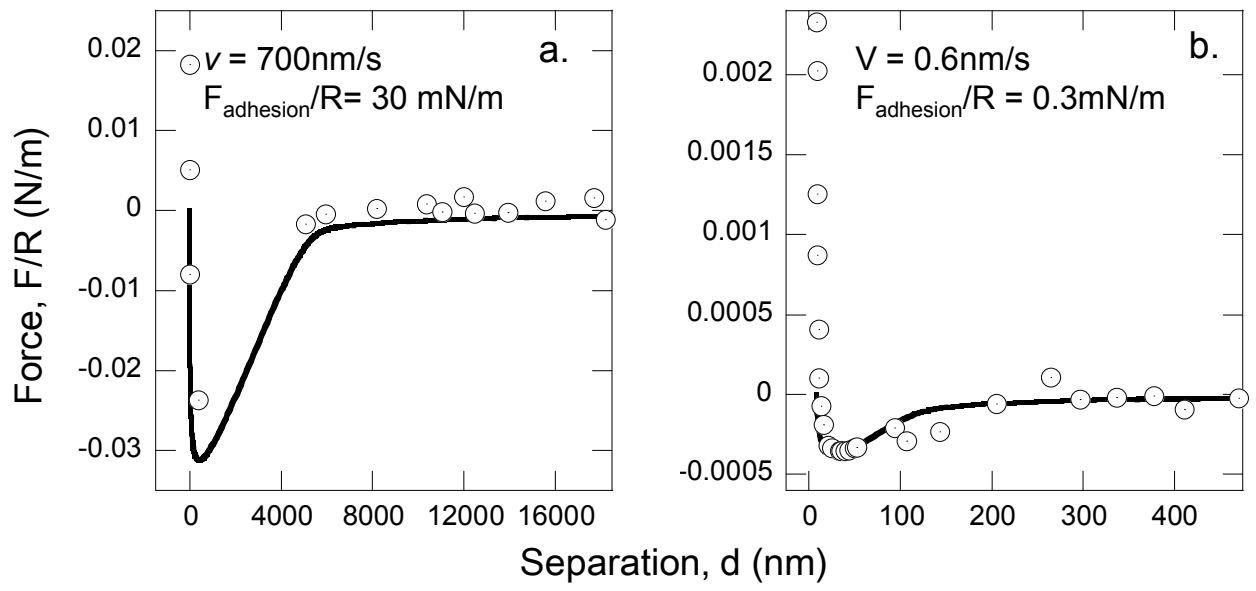

Figure 2. Attractive viscous forces measured when pulling a PMMA surface in contact with a silver film measured for two different drive velocities.

## Conclusions

Hydrodnamic drainage forces were measured between PMMA and silver in silicon oil. The measured forces are in good agreement with Reynolds' theory and will later be compared to squeeze out forces between structured surfaces.

## Acknowledgements

This work is supported by the National Science Foundation under Grant CMMI-0709187, and 3M Corp.

## References

1. K. Autumn, *MRS Bull.*, 2007, 32, pp 473–478.
2. K. Autumn, Y. Liang, S. Hsieh, W. Zesch, *et al.*, *Nature*, 2000, 405, pp 681-685.
3. K. Autumn, M. Sitti, Y. Liang, A. Peattie, *et al.*, *Proc. Nat. Acad. Sci.*, 2002, 99, pp 12252.
4. W. Hansen and K. Autumn, *Proc. Nat. Acad. Sci.*, 2005, 102, pp 385.
5. N. Pesika, Y. Tian, B. Zhao, K. Rosenberg, *et al.*, *J. Adhesion*, 2007, 83, pp 383-401.
6. Y. Tian, N. Pesika, H. Zeng, K. Rosenberg, *et al.*, *Proc. Nat. Acad. Sci.*, 2006, 103, pp 19320.
7. E. Arzt, S. Gorb and R. Spolenak, *Proc. Nat. Acad. Sci.*, 2003, 100, pp 10603.
8. W. B. Russel, D. A. Saville and W. R. Schowalter *Colloidal Dispersions*; Cambridge University Press: New York, 1989.
9. J. N. Israelachvili, *J. Vac. Sci. Tech. A*, 1991, 10, pp 2961-2971.
10. W. Federle, W. Barnes, W. Baumgartner, P. Drechsler, *et al.*, *J. R. Soc. Interface*, 2006, 3, pp 689.
11. J. N. Israelachvili, *J. Colloid Interface Sci.*, 1973, 44, pp 259-272.
12. D. Tabor and R. H. S. Winterton, *Proc. R. Soc. London Ser A-Math.*, 1969, 312, pp 435-450.
13. P. M. McGuiggan, J. Zhang and S. M. Hsu, *Tribol. Lett.*, 2001, 10, pp 217-223.
14. S. Tolansky *Multiple-Beam Interferometry of surfaces and films*; Oxford University Press: London, 1948.
15. O. Reynolds, *Philos. Trans. R. Soc. Lond.*, 1886, 177, pp 157.
16. D. Y. C. Chan and R. G. Horn, *J. Chem. Phys.*, 1985, 10, pp 5311-5324.